\documentclass[pre,twocolumn,superscriptaddress]{revtex4-1}

\pdfoutput=1

\usepackage{graphicx}
\usepackage[caption=false]{subfig}
\usepackage{amsmath,amssymb,gensymb}
\usepackage{color}
\usepackage{mathtools}

\usepackage{upgreek}
\usepackage{bm}

\newcommand{\am}[1]{{\textcolor{black}{#1}}}

\begin{document}

\title{Memory-based mediated interactions between rigid particulate inclusions\\ in viscoelastic environments}

\author{Mate Puljiz}
\email{puljiz@thphy.uni-duesseldorf.de}
\affiliation{Institut f{\"u}r Theoretische Physik II: Weiche Materie, 
Heinrich-Heine-Universit{\"a}t D{\"u}sseldorf, D-40225 D{\"u}sseldorf, Germany}
\author{Andreas M. Menzel}
\email{menzel@thphy.uni-duesseldorf.de}
\affiliation{Institut f{\"u}r Theoretische Physik II: Weiche Materie, 
Heinrich-Heine-Universit{\"a}t D{\"u}sseldorf, D-40225 D{\"u}sseldorf, Germany}

\date{\today}

\begin{abstract}
Many practically relevant materials combine properties of viscous fluids and elastic solids to viscoelastic behavior. \am{Our focus is on the induced dynamic behavior of damped finite-sized particulate inclusions in such substances. We explicitly describe history-dependent interactions that emerge between the embedded particles.  These interactions are mediated by the viscoelastic surroundings. They result from the flows and distortions of the viscoelastic medium when induced by the rigid inclusions. Both, viscoelastic environments of terminal fluid-like flow or of completely reversible damped elastic behavior, are covered. For illustration and to highlight the role of the formalism in potential applications, we briefly address the relevant examples of dragging a rigid sphere through a viscoelastic environment together with subsequent relaxation dynamics, the switching dynamics of magnetic fillers in elastic gel matrices, and the swimming behavior of active microswimmers in viscoelastic solutions. The approach provides a basis for more quantitative and extended investigations of these and related systems in the future.} 
\end{abstract}

\maketitle

\section{Introduction}


%
%
%
%

To a large extent, real materials do not behave in a purely liquid- or solid-like way. Substances that on long time scales show a fluid-like terminal flow may feature reversible quasi-elastic deformations under sufficiently short application of external forces. Prominent examples 
are melts or solutions of 
entangled polymer molecules \cite{gennes1979scaling,doi1988theory}. Vice versa, the dynamics of materials that do feature reversible elastic deformations on long time scales, like rubbers or soft elastic polymeric gels, may still be damped by internal viscous-like friction 
\cite{gennes1979scaling,doi1988theory}. 

Materials 
combining such viscous 
and elastic 
characteristics are termed viscoelastic. Studying their behavior on a mesoscopic particulate length scale becomes important when they contain immersed 
or embedded rigid inclusions. Especially, this concerns interactions between the embedded particles 
mediated by the viscoelastic environment. 
They arise 
when forces or torques 
imposed on 
or 
actively generated by the particles are transmitted to their 
surroundings. 
Recently studied 
experimental examples 
comprise: microrheological investigations probing the 
environment 
of embedded colloidal particles 
when driving them via external 
magnetic or optical fields 
\cite{ziemann1994local,waigh2005microrheology,wilhelm2008out, bausch1999measurement,wilson2011small}; 
switching the magnetic interactions between rigid inclusions in soft elastic gel matrices by external magnetic fields \cite{filipcsei2007magnetic, an2014conformational,huang2016buckling, puljiz2016forces}; or self-driven active microswimmers \cite{elgeti2015physics,bechinger2016active,zottl2016emergent, menzel2015tuned} 
propelling through viscoelastic environments \cite{gaffney2011mammalian,li2016collective, gomez2016dynamics,lozano2018run, datt2017active}. 

Here, we provide a corresponding theoretical framework. 
\am{Our focus is on systems featuring discrete finite-sized particulate inclusions embedded in a surrounding viscoelastic continuous matrix. We concentrate on the overdamped dynamics of the embedded particles. 
The interactions between the embedded particles that are mediated by the viscoelastic environment are covered. Moreover,} our approach 
allows to describe both long-time terminal flow 
and long-time reversible elastic behavior of the viscoelastic surroundings. 
Accordingly, 
the 
broad range of physical systems mentioned above can be addressed. 
Later, 
our results will 
serve to study, e.g., by statistical means, links between the collective behavior of 
embedded particle ensembles and overall material properties \cite{arteinstein1906,arteinstein1911,smallwood1944limiting, batchelor1972determination,dhont1996introduction, morris1996self,rex2008dynamical,cremer2017density, menzel2016dynamical}. 

\am{Next, we introduce in Sec.~\ref{matrix} the continuum description that we use to represent the viscoelastic environment. The linearity of the employed approach allows to derive and use the Green's function associated with a localized force impulse acting within the viscoelastic medium. We proceed in Sec.~\ref{inclusions} by inserting and quantifying the role of finite-sized rigid spherical inclusions in this viscoelastic background. To illustrate in Sec.~\ref{examples} the strength of the approach, we then briefly and qualitatively address its possible application to different examples. This opens the way to a broad range of more detailed and quantitative studies in the future. Some conclusions are listed in Sec.~\ref{conclusions}. Apart from that, we illustrate in App.~\ref{AppHydro} how the continuum equation characterizing our viscoelastic environment can be obtained from a generalized hydrodynamic approach \cite{temmen2000convective}. Moreover, we include in App.~\ref{AppGreen} the detailed derivation of the Green's function mentioned above. In App.~\ref{AppKramers}, we demonstrate that this Green's function satisfies the Kramers--Kronig relations as required. A basic example situation considered in App.~\ref{AppMatEl} serves to further illustrate our approach. Finally, we show in App.~\ref{AppTorque} that the formalism used here to address the role of net forces acting on rigid particles embedded in a viscoelastic environment is readily extended to include net torques as well.}

\section{Basic continuum description of the viscoelastic matrix}\label{matrix}

To characterize the behavior of the viscoelastic environment, we consider an isotropic and homogeneous 
viscoelastic continuum on the scale of the mesoscopic inclusions. 
A simple linearized description will be employed to allow for an effective analytical treatment.
First, we recall the 
limits of purely viscous and purely linearly elastic materials. 
Incompressible viscous low-Reynolds-number fluid flows 
are quantified by Stokes's equations \cite{stokes1845theories,dhont1996introduction}
\begin{equation}\label{Stokes}
	\eta\nabla^2 \mathbf{v}(\mathbf{r},t) = \nabla p(\mathbf{r},t) -\mathbf{f}_b(\mathbf{r},t),
	\qquad
	\nabla\cdot\mathbf{v}(\mathbf{r},t) = 0,
\end{equation}
with $\eta$ the dynamic viscosity, $\mathbf{v}(\mathbf{r},t)$ the flow velocity field, 
$p(\mathbf{r},t)$ the pressure field, and $\mathbf{f}_b(\mathbf{r},t)$ the bulk force density acting on the fluid. 
Contrariwise, reversible deformations of incompressible linearly elastic solids are quantified by the Navier--Cauchy equations \cite{cauchy1828sur}
\begin{equation}\label{NC}
	\mu\nabla^2 \mathbf{u}(\mathbf{r},t) {}={} -\mathbf{f}_b(\mathbf{r},t), 
	\qquad
	\nabla\cdot\mathbf{u}(\mathbf{r},t) =  0,
\end{equation}
with $\mu$ the shear modulus, $\mathbf{u}(\mathbf{r},t)$ the displacement field of the material elements, 
and $\mathbf{f}_b(\mathbf{r},t)$ the bulk force density acting on the 
solid. 
\am{The assumed incompressibility represents a reasonable approximation in many cases. Namely, this should include (semi)dilute aqueous solutions of polymers \cite{gomez2016dynamics,narinder2018memory} and classes of swollen polymeric gels \cite{puljiz2016forces,puljiz2018reversible}.}  

For an incompressible, infinitely extended, isotropic, and homogeneous viscoelastic medium, 
we now combine these balances of force densities, 
Eqs.~(\ref{Stokes}) and (\ref{NC}), yielding 
\begin{equation}\label{visco}
	\mu\nabla^2 \mathbf{u}(\mathbf{r},t) + \eta \nabla^2\mathbf{v}(\mathbf{r},t) ={} \nabla p(\mathbf{r},t) -\mathbf{f}_b(\mathbf{r},t).
\end{equation}
This relation can be confirmed by 
linearizing a general continuum approach 
based purely on conservation laws and symmetry arguments \cite{temmen2000convective}, \am{constricted to the regime of the overdamped dynamics considered here. See App.~\ref{AppHydro} for the details.} 
\am{Moreover, in this limit, Eq.~(\ref{visco}) likewise follows when combining the dynamic equations of a two-fluid approach that contains frictional coupling between an elastic component and a viscous fluid component \cite{levine2000one,levine2001response,yasuda2018three}.}


Formally, linear elasticity theory does not 
distinguish whether the coordinates $\mathbf{r}$ 
refer to the initial (undeformed) or the present 
(deformed) state of a material \cite{landau1986theory,temmen2000convective}. Yet, 
hydrodynamics 
dictates the latter, Euler point of view \cite{temmen2000convective}. 
Thus, 
$\mathbf{u}(\mathbf{r},t)$ 
describes the reversible elastic displacements that have taken the material elements \textit{to} the positions $\mathbf{r}$. 
\am{Or, probably more appropriately in the present context, $\mathbf{u}(\mathbf{r},t)$ quantifies the \textit{memory} of those positions that the material elements would tend to displace back to from their present positions, if the material is relaxed from its current state.} 

In a viscoelastic medium, this memory of the initial positions stored in $\mathbf{u}(\mathbf{r},t)$ may fade away over time. In addition to 
changes in $\mathbf{u}(\mathbf{r},t)$ 
arising from material motion $\mathbf{v}(\mathbf{r},t)$, 
we therefore assume a simple relaxation
\begin{equation}\label{velocity_field}
	\mathbf{\dot{u}}(\mathbf{r},t) ={} \mathbf{v}(\mathbf{r},t)-\gamma \mathbf{u}(\mathbf{r},t). 
\end{equation}
The relaxation rate $\gamma$ sets the ``forgetfulness'' of the medium, implying complete reversibility for $\gamma=0$. 
\am{(This equation is different from the corresponding one in Refs.~\onlinecite{levine2000one,levine2001response,yasuda2018three}, where instead of the relaxational decay a diffusional contribution appears).}
Eliminating $\mathbf{v}(\mathbf{r},t)$ from Eqs.~(\ref{visco}) and (\ref{velocity_field}), we obtain
\begin{equation}\label{Combined2}
	(\mu+\gamma\eta)\nabla^2 \mathbf{u}(\mathbf{r},t) + \eta \nabla^2\mathbf{\dot{u}}(\mathbf{r},t) ={} \nabla p(\mathbf{r},t) -\mathbf{f}_b(\mathbf{r},t). 
\end{equation}

Due to the linearity in $\mathbf{u}(\mathbf{r},t)$, the 
Green's function can be determined.  
It solves Eq.~(\ref{Combined2}) 
for a given force impact $\mathbf{F}$ at position $\mathbf{R}_0$ and time $t_0$, setting $\mathbf{f}_b(\mathbf{r},t)=\mathbf{F}\delta(\mathbf{r}-\mathbf{R}_0)\delta(t-t_0)$. 
The procedure is well-established, 
e.g., in low-Reynolds-number hydrodynamics \cite{karrila1991microhydrodynamics, dhont1996introduction} or linear elasticity theory \cite{teodosiu1982elastic,puljiz2016forces,puljiz2017forces}. After transforming to Fourier space, here both in space and time, solving for the displacement field, 
projecting on transverse modes according to the incompressibility relation in Eq.~(\ref{NC}), and transforming back, we obtain 
\begin{equation}\label{green_definition}
	\mathbf{u}(\mathbf{r},t) ={} \mathbf{\underline{G}}(\mathbf{r}-\mathbf{R}_0,t-t_0)\cdot\mathbf{F},
\end{equation}
with $\mathbf{\underline{G}}(\mathbf{r},t)$ the second-rank tensor
\begin{equation}\label{greens_function_visco}
	\mathbf{\underline{\hspace{-.02cm}G}}(\mathbf{r}, t) ={}  \frac{1}{8\pi\eta |\mathbf{r}|} \left[\mathbf{\underline{\hat{I}}}+\mathbf{\hat{r}}\mathbf{\hat{r}}\right]\Theta(t)\mathrm{e}^{-\frac{\mu+\gamma\eta}{\eta}t}.
\end{equation}
$\mathbf{\underline{\hat{I}}}$ is the identity matrix, $\mathbf{\hat{r}}\mathbf{\hat{r}}$ a dyadic product with $\mathbf{\hat{r}}=\mathbf{r}/|\mathbf{r}|$, and $\Theta(t)$ the Heaviside function. (For readers unfamiliar with this common 
technique, we reproduce it in App.~\ref{AppGreen}, together with a remark on the associated Kramers--Kronig relations in App.~\ref{AppKramers}). We note that $\mathbf{\underline{\hspace{-0.02cm}G}}(\mathbf{r},t)=\mathbf{\underline{\hspace{-0.02cm}G}}(\mathbf{r})G(t)$, with  
\begin{equation}
	\mathbf{\underline{\hspace{-.02cm}G}}(\mathbf{r})=  \frac{1}{8\pi\eta |\mathbf{r}|}\left[\mathbf{\underline{\hat{I}}}+\mathbf{\hat{r}}\mathbf{\hat{r}}\right], \qquad
	G(t)= \Theta(t)\mathrm{e}^{- \frac{\mu+\gamma\eta}{\eta}t}, \label{Greens_function_split}
\end{equation}
where $\mathbf{\underline{G}}(\mathbf{r})$ has the same form as the hydrodynamic Oseen tensor \cite{dhont1996introduction}.
The viscoelastic displacements resulting from a general force density $\mathbf{f}_b(\mathbf{r},t)$ are thus obtained as
\begin{equation}\label{displacement-force-field}
	\mathbf{u}(\mathbf{r},t) =
	\int_{\mathbb{R}}\mathrm{d}t'~G(t-t')
	\int_{\mathbb{R}^3} \mathrm{d}^3r'~
	 \mathbf{\underline{G}}(\mathbf{r}-\mathbf{r}')\cdot\mathbf{f}_b(\mathbf{r}',t').
\end{equation}
\am{For illustration, we consider the example trajectory of one material element subject to an interim constant and otherwise vanishing concentrated force density in App.~\ref{AppMatEl}.}

\section{Rigid spherical inclusions}\label{inclusions}

We now turn to rigid spherical particles of radius $a$ embedded in the viscoelastic medium. No-slip conditions prevail on their surfaces. 
An external force $\mathbf{F}(t)$ exerted on a particle centered at $\mathbf{R}(t)$ 
is transmitted to the environment, 
distorting it, and/or setting it into motion. 
Starting from the Green's function and 
the formal analogy 
of $\mathbf{\underline{G}}(\mathbf{r})$ to 
the hydrodynamic case \cite{karrila1991microhydrodynamics}, we consider the displacement field 
%
%
\begin{eqnarray}\label{u_general}
	\mathbf{u}(\mathbf{r},t)
	&=&
	\int_\mathbb{R}\!\mathrm{d}t'\,G(t-t')\:\Bigg\{ \!
    \nonumber\\&&{} 
		\left[\left(1+\frac{a^2}{6}\nabla^2\right)	  
		\mathbf{\underline{G}}\big(\mathbf{s}(t')
		\big)\right]\! \cdot\mathbf{F}(t') 
		\,\Theta\big(|\mathbf{s}(t')|-a
				\big)	
		\nonumber\\
		&&{}
		+\frac{1}{6\pi\eta a}\mathbf{F}(t')~\Theta\big(a-|\mathbf{s}(t')|
						\big)
		\Bigg\},	\qquad
\end{eqnarray}
%
with $\mathbf{s}(t'):=\mathbf{r}-\mathbf{R}(t')$
and a continuous integrand for each $|\mathbf{s}(t')|=a$. 
This expression solves inside the embedding medium, 
for $|\mathbf{s}(t)|>a$, the linear Eq.~(\ref{Combined2}) for $\mathbf{f}_b(\mathbf{r},t)=\mathbf{0}$, together with the pressure field $p(\mathbf{r},t)=\mathbf{s}(t)/4\pi\left|\mathbf{s}(t)\right|^3\cdot\mathbf{F}(t)$ as in the hydrodynamic case \cite{karrila1991microhydrodynamics}. 
It satisfies $\nabla\cdot\mathbf{u}(\mathbf{r},t)=0$ and the boundary condition $\mathbf{u}(\mathbf{r},t)\rightarrow\mathbf{0}$ for $|\mathbf{s}(t)|\rightarrow\infty$, see Eq.~(\ref{greens_function_visco}), if all $|\mathbf{R}(t')|$ remain finite. 
At each instant in time $t'$, the expression in the square brackets is constant 
on the surface of the sphere 
($|\mathbf{s}(t')|=a$), 
reflecting its rigid 
displacement 
and confirmed by explicitly evaluating $(1+a^2\nabla^2/6)\mathbf{\underline{G}}\big(\mathbf{r}-\mathbf{R}(t')\big)$. 
Similar relations can be obtained when applying a net torque, see App.~\ref{AppTorque}.  
We determine the 
reaction of the sphere at time $t$ to the contributions generated by itself at earlier times $t'<t$, see Eq.~(\ref{u_general}), 
via Eqs.~(\ref{u_integral}) and (\ref{omega_integral}) below.

At each point in time $t$, the no-slip 
condition on the particle surface $\mathbf{r}\in\partial V$ reads
%
%
\begin{equation}\label{faxen_boundary}
	{\mathbf{U}(t)+\boldsymbol{\Omega}(t)\times\big(\mathbf{r}-\mathbf{R}(t)\big) }
	= \mathbf{u}(\mathbf{r}, t). 
\end{equation}
%
It states that the displacements of the particle surface points, 
given by its rigid displacement $\mathbf{U}(t)$ and rotation $\boldsymbol{\Omega}(t)$, must be equal to the displacement $\mathbf{u}(\mathbf{r},t)$ of the there anchored surrounding medium. 
Here, $\mathbf{u}(\mathbf{r},t)$ contains the displacements in the medium induced by the particle itself, also at earlier times, see Eq.~(\ref{u_general}), and all displacements generated by all other sources. 
Due to the 
linearity of Eq.~(\ref{Combined2}), all contributions 
superimpose.  

Equation~(\ref{faxen_boundary}) allows to determine 
at each time $t$ how a rigid sphere is displaced [$\mathbf{U}(t)$] and rotated [$\boldsymbol{\Omega}(t)$] in a given displacement field $\mathbf{u}(\mathbf{r},t)$. The derivation of these so-called Fax\'en laws follows the same lines as in low-Reynolds-number hydrodynamics \cite{dhont1996introduction, karrila1991microhydrodynamics} and linear elasticity theory \cite{kim1994faxen,norris2008faxen, puljiz2016forces,puljiz2017forces}. 
%
%
Integrating both sides of Eq.~(\ref{faxen_boundary}) 
over 
the surface $\partial V$ of the sphere, 
the antisymmetric $\boldsymbol{\Omega}(t)$-term vanishes and we obtain
\begin{equation}\label{u_integral}
	\mathbf{U}(t) ={} \frac{1}{4\pi a^2}\int_{\partial V}\mathrm{d}^2 |\mathbf{r}-\mathbf{R}(t)|~ \mathbf{u}(\mathbf{r},t).
\end{equation}
Similarly, 
multiplying Eq.~(\ref{faxen_boundary}) 
by $\big(\mathbf{r}-\mathbf{R}(t)\big)\times$ before the integration, 
we find
\begin{equation}\label{omega_integral}
	\boldsymbol{\Omega}(t) ={} \frac{3}{8\pi a^4}\int_{\partial V}\mathrm{d}^2 |\mathbf{r}-\mathbf{R}(t)|~\big(\mathbf{r}-\mathbf{R}(t)\big)\times \mathbf{u}(\mathbf{r},t).
\end{equation}
These integrals can be evaluated numerically 
\cite{beentjes2015quadrature}. 
If $\mathbf{u}(\mathbf{r},t)$ in Eqs.~(\ref{u_integral}) and (\ref{omega_integral}) is infinitely differentiable, we may expand it in $\mathbf{s}(t)=\mathbf{r}-\mathbf{R}(t)$ as 
$\mathbf{u}(\mathbf{R}(t)+\mathbf{s}(t),t)=(1+\mathbf{s}(t)\cdot\nabla	
	+\mathbf{s}(t)\mathbf{s}(t)\!:\!\nabla\nabla/2+\dots)\mathbf{u}(\mathbf{r},t)|_{\mathbf{r}=\mathbf{R}(t)}$. 
Uneven dyadics 
in $\mathbf{s}(t)$ vanish upon integration $\int_{\partial V}\mathrm{d}^2|\mathbf{s}(t)|$, 
even dyadics of $\mathbf{s}(t)$ lead to dyadic combinations of $\mathbf{\underline{\hat{I}}}$, and 
$\nabla^{(2n)}\mathbf{u}(\mathbf{r},t)=\mathbf{0}=\nabla\times\nabla^2\mathbf{u}(\mathbf{r},t)$ ($n\geq2$ integer), which 
follows from directly applying these differential operators to the Green's function in Eq.~(\ref{greens_function_visco}). Then,  
the Fax\'en relations are again of the form \cite{dhont1996introduction, karrila1991microhydrodynamics,kim1994faxen,norris2008faxen, puljiz2016forces,puljiz2017forces}
%
%
%
%
%
%
%
%
\begin{eqnarray}
    \label{faxen_translation}
	\mathbf{U}(t) &=& \left(1+\frac{a^2}{6}\nabla^2\right)\mathbf{u}(\mathbf{r},t)\bigg|_{\mathbf{r}=\mathbf{R}(t)}, \\
%
%
%
    \label{faxen_rotation}
	\boldsymbol{\Omega}(t) &=& \frac{1}{2}\nabla\times\mathbf{u}(\mathbf{r},t)\bigg|_{\mathbf{r}=\mathbf{R}(t)}.
\end{eqnarray}
%
%
%

Correspondingly, the analog to Eq.~(\ref{faxen_boundary}) links the velocity $\mathbf{V}(t)$ and angular velocity $\mathbf{W}(t)$ of the sphere to the velocity field $\mathbf{v}(\mathbf{r},t)$ of the environment. 
%
%
Repeating the above steps, we obtain the 
associated Fax\'en laws in analogy to Eqs.~(\ref{u_integral})--(\ref{faxen_rotation}), simply 
replacing $\mathbf{U}(t)\rightarrow\mathbf{V}(t)$, $\boldsymbol{\Omega}(t)\rightarrow\mathbf{W}(t)$, and $\mathbf{u}(\mathbf{r},t)\rightarrow\mathbf{v}(\mathbf{r},t)$. 
%
%
These relations are consistent with Eq.~(\ref{velocity_field}), 
%
%
$\mathbf{V}(t) = \mathbf{\dot{U}}(t)+\gamma\mathbf{U}(t)$,
and
$\mathbf{W}(t) = \boldsymbol{\dot{\Omega}}(t)+\gamma\boldsymbol{\Omega}(t)$. 

\am{We stress the physical importance of the velocity field and the particle velocities. If one is interested in the local material transport within the material, the important quantity is the velocity field $\mathbf{v}(\mathbf{r},t)$, and not predominantly the \textit{memory} displacement field $\mathbf{u}(\mathbf{r},t)$. The latter at each instant in time describes towards where the material points would tend to relax, if all forces exerted on the material are switched off. For $\gamma\neq0$, these in general are not the positions that the material points had started from initially because the memory decays over time. In our case, we consider the transport of the rigid spheres. Their total net translation is given by integration of their velocities $\mathbf{V}(t)$ over their course of motion, and not simply by the value of the current memory variables $\mathbf{U}(t)$.}

Accordingly, 
our picture is closed. 
From Eq.~(\ref{u_general}), we 
calculate $\mathbf{u} (\mathbf{r},t)$ at each requested position $\mathbf{r}$ and time $t$, resulting from the forces exerted by the spherical particles on their viscoelastic environment. 
$\mathbf{v}(\mathbf{r},t)$ follows via Eq.~(\ref{velocity_field}). From the Fax\'en relations we obtain the velocities $\mathbf{V}(t)$ and angular velocities $\mathbf{W}(t)$ of the particles. Integrating these over time leads to the particle trajectories $\mathbf{R}(t)$ and courses of rotation. 

What we neglect in this approach are additional contributions to the displacement field that arise from the resistance of the rigid particles to their deformations under strong local distortion \cite{puljiz2016forces}. 
In the hydrodynamic language, we here stop at the common Rodne-Prager level \cite{dhont1996introduction}, rendering the results qualitative 
when our spheres meet areas of stronger distortion. 
%
Moreover, our formulae are evaluated numerically. 
By construction, this involves finite time steps and finite displacements. 
Technically, to apply in our numerical discretization the force density $\mathbf{f}_b(\mathbf{r},t)$ in Eq.~(\ref{Combined2}) at the positions where it causes the resulting displacements, 
we shift back the force centers at each time step according to the memorized displacements to
calculate the induced $\mathbf{u}(\mathbf{r},t)$. 
%
This scheme correctly reproduces 
complete elastic reversibility for $\gamma\rightarrow0$ \cite{suppl}. 

\section{Examples}\label{examples}

\am{To demonstrate the range of the theory and for illustration, we now briefly consider 
several examples. 
First, this concerns a single sphere dragged by a net force through a viscoelastic environment. In this context, we also demonstrate that the familiar Green's functions associated with incompressible linearly elastic solids and low-Reynolds-number incompressible fluid flows are reproduced 
in the corresponding limits. Next, the pairwise interactions between magnetizable finite-sized particles are considered. These particles interact both magnetically and via the flows and distortions induced in their viscoelastic surroundings. Afterwards, we touch the topic of self and mutual interactions of active self-propelled microswimmers in viscoelastic media.}

\subsection{Dragged rigid sphere}\label{draggedsphere}

\begin{figure}
\centerline{\includegraphics[width=\columnwidth]{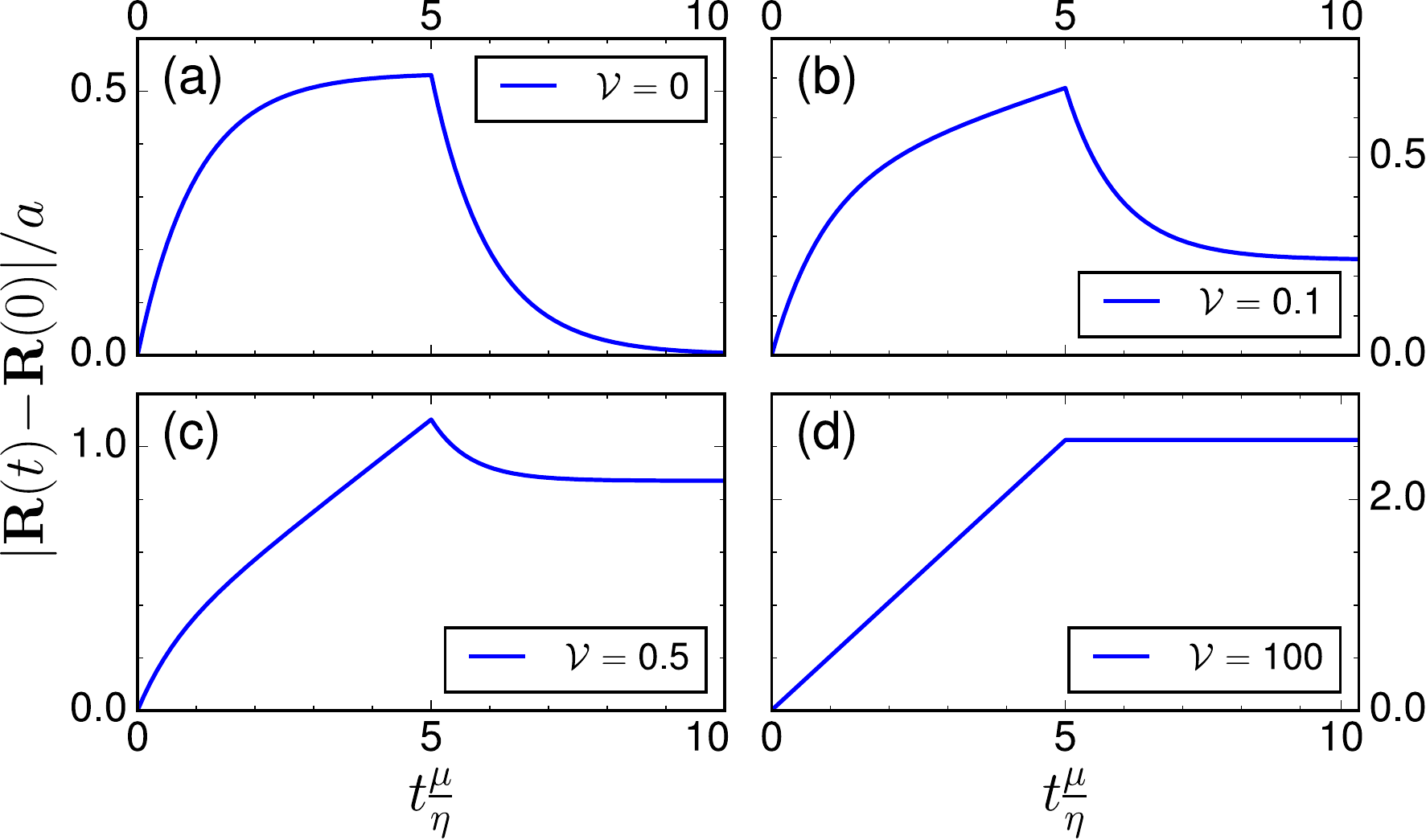}}
\caption{
Overall translation $|\mathbf{R}(t)-\mathbf{R}(0)|$ 
for a spherical particle of radius $a$, 
dragged by a constant force of magnitude $|\mathbf{F}|=10\mu a^2$ 
during times $0\leq t\leq 5\eta/\mu$. 
$\mathcal{V}={\gamma\eta}/{\mu}$ 
sets the memory of the viscoelastic environment 
and determines 
the degree of relaxation back to the initial position for $t>5\eta/\mu$, when the drag force has been switched off again.
(a) $\mathcal{V}=0$ 
implies fully reversible elasticity, while  
(b,c) intermediate values of $\mathcal{V}$ 
imply partial loss of memory and partial reversibility. 
(d) 
$\mathcal{V}=100$ already 
shows the phenomenology of viscous hydrodynamics ($\mathcal{V}\rightarrow\infty$). 
\am{The ordinates show different scales, so that the magnitude of maximum translation grows from (a) to (d).} 
}
\label{fig_1sphere_constant_force}
\end{figure}

Figure~\ref{fig_1sphere_constant_force} starts with the basic scenario of dragging one spherical particle through a viscoelastic medium by a constant external force, 
switched on and off at given times. 
For instance, this situation concerns active magnetic microrheological measurements, where colloidal particles are driven by external magnetic field gradients to probe the viscoelastic 
environment \cite{ziemann1994local,waigh2005microrheology,wilhelm2008out, bausch1999measurement,wilson2011small}. 
%
%

The description correctly reproduces the limiting cases.  
To confirm this, we meanwhile measure 
lengths in particle radii $a$, 
bulk force densities in units of $\mu/a$, and time in units of $\eta/\mu$. 
Then, the whole system behavior is controlled by 
one \am{remaining}
dimensionless number 
\begin{equation}\label{visco_parameter}
	\am{\mathcal{V} ={} \frac{\gamma\eta}{\mu}}    
\end{equation}
\am{appearing} 
in the exponent in Eqs.~(\ref{greens_function_visco}) or (\ref{Greens_function_split}), \am{and particularly as the relaxation parameter in Eq.~(\ref{velocity_field}).}  

Indeed, for $\mathcal{V}\rightarrow0$, complete elastic reversibility is recovered, see Fig.~\ref{fig_1sphere_constant_force}(a). Theoretically, this 
follows from switching on a static force density $\mathbf{f}_b(\mathbf{r})$ at $t=t_0$ in Eq.~(\ref{displacement-force-field}) via $\mathbf{f}_b(\mathbf{r},t)=\mathbf{f}_b(\mathbf{r})\Theta(t-t_0)$. After complete relaxation for $t\rightarrow\infty$, we obtain the correct steady-state displacements in an incompressible linearly elastic solid 
\cite{teodosiu1982elastic,phan1993rigid,   puljiz2016forces,puljiz2017forces} 
\begin{equation}
	\mathbf{u}(\mathbf{r},t) = \int_{\mathbb{R}^3}\mathrm{d}^3 r'~\frac{\eta}{\mu}\mathbf{\underline{G}}(\mathbf{r}-\mathbf{r}')\cdot\mathbf{f}_b(\mathbf{r}').
\end{equation}
Conversely, 
$\mathcal{V}\gg1$ implies the hydrodynamic limit of negligible memory, see Fig.~\ref{fig_1sphere_constant_force}(d). In the theory, 
we take the time derivative of Eq.~(\ref{displacement-force-field}), insert it as ${\mathbf{\dot{u}}}(\mathbf{r},t)$ into Eq.~(\ref{velocity_field}), and then take the limit $\mathcal{V}\rightarrow\infty$ of vanishing elasticity. This leads to
\begin{equation}
	\mathbf{v}(\mathbf{r},t) = \int_{\mathbb{R}^3}\mathrm{d}^3 r'~\mathbf{\underline{G}}(\mathbf{r}-\mathbf{r}')\cdot\mathbf{f}_b(\mathbf{r}',t),
\end{equation}
reproducing the correct Oseen expression of low-Reynolds-number hydrodynamics 
\cite{karrila1991microhydrodynamics,dhont1996introduction}. 
Intermediate \am{finite values of $\mathcal{V}>0$ imply viscoelasticity} of partially decaying memory, \am{with example cases} depicted in Fig.~\ref{fig_1sphere_constant_force}(b) and (c). 



\subsection{Magnetically induced particle interactions}

A more specific example 
of practical relevance is given by magnetic hybrid materials 
of magnetic or magnetizable colloidal particles embedded in soft polymeric environments \cite{ filipcsei2007magnetic,menzel2015tuned, odenbach2016microstructure,weeber2018polymer}. 
Manipulating the magnetic particle interactions 
by external magnetic fields 
allows to reversibly 
tune from outside the mechanical material stiffness 
\cite{filipcsei2007magnetic,schumann2017situ}. Studying the switching dynamics on the particle scale \cite{roeder2015magnetic,ilg2018magnetic,goh2018dynamics, huang2016buckling,puljiz2016forces} becomes especially important in applications as 
soft actuators \cite{filipcsei2007magnetic, fuhrer2009crosslinking,ilg2013stimuli,thevenot2013magnetic, li2011design}. 

\begin{figure*}
\centerline{\includegraphics[width=1.75\columnwidth]{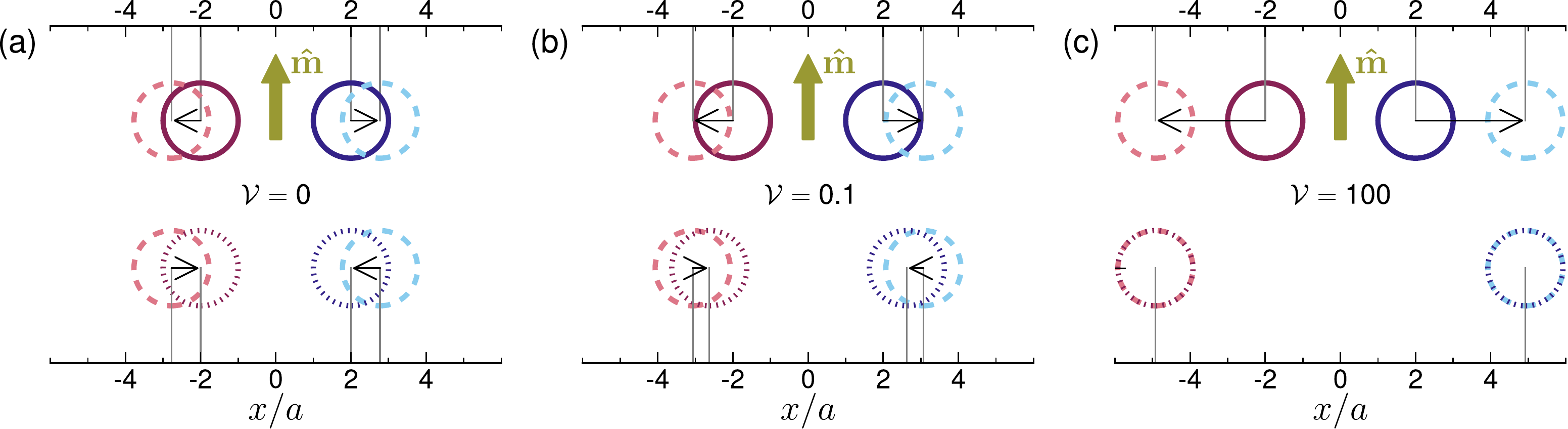}}
\caption{Two magnetizable spherical particles 
are exposed to a saturating external magnetic field. 
Repulsive magnetic dipole moments $\mathbf{m}=300(\mu a^6/\mu_0)^{1/2}\mathbf{\hat{m}}$ 
are induced during times $0\leq t\leq 10\eta/\mu$ (upper row) and then switched off again 
(bottom row), 
for (a) complete elastic reversibility $\mathcal{V}=0$, (b) $\mathcal{V}=0.1$, and (c) basically absent memory $\mathcal{V}=100$.
%
%
Circles mark the positions at $t=0$ (solid), $t=10\eta/\mu$ (dashed), and $t\gg10\eta/\mu$ (dotted). 
%
Black arrows indicate 
trajectories.
}
\label{fig_mag_parallel}
\end{figure*}

Approximating the magnetic moments of the particles by 
dipoles \cite{klapp2005dipolar}, we consider for illustration the effect of 
pairwise 
interactions between two identical spherical paramagnetic particles, labeled 
$1$ and $2$. For simplicity, a strong homogeneous external magnetic field is applied, leading to identical saturated magnetic 
moments $\mathbf{{m}}$. 
The dipolar 
interaction force on particle $i$ is given by \cite{jackson1962classical}
\begin{equation}\label{magnetic_force}
	\mathbf{F}_i(t) ={} \frac{3\mu_0|\mathbf{m}|^2}{4\pi} \frac{2\mathbf{\hat{m}}
	\big(
			\mathbf{\hat{m}}\hspace{-1pt}\cdot\hspace{-1pt}\mathbf{\hat{d}}(t)
		\big)+
		\mathbf{\hat{d}}(t)-5\mathbf{\hat{d}}(t)\big(
		\mathbf{\hat{m}}\hspace{-1pt}\cdot\hspace{-1pt}\mathbf{\hat{d}}(t)
	\big)^2
	}{|\mathbf{d}(t)|^4},
\end{equation}
with $\mu_0$ the magnetic vacuum permeability, $\mathbf{\hat{m}}=\mathbf{m}/|\mathbf{m}|$, $\mathbf{d}(t)=\mathbf{R}_i(t)-\mathbf{R}_{j\neq i}(t)$, and  $\mathbf{\hat{d}}(t)=\mathbf{d}(t)/|\mathbf{d}(t)|$ ($i,j\in\{1,2\}$). 

%
%
\am{In Fig.~\ref{fig_mag_parallel} we consider two such magnetizable spherical particles embedded in different viscoelastic background media. The solid circles indicate the initial positions in the unmagnetized state.} 
\am{Next, the particles are magnetized by a strong external magnetic field in a way that they repel each other. After a longer time of magnetization, the positions in the fully reversible elastic case (a) of $\mathcal{V}=0$ are set by the balance of magnetic repulsive and counteracting mechanical restoring forces, the latter resulting from the elastic distortion. In the other cases (b) and (c) of $\mathcal{V}\neq0$, an in principle unbounded withdrawal of the particles from each other is observed with elapsing time. Yet, the speed of withdrawal decreases as the magnetic repulsion drops with increasing mutual particle distance. Dashed circles mark the positions attained when the field is switched off again.} 

\am{The second row of Fig.~\ref{fig_mag_parallel} then shows the reaction after switching off the induced magnetic repulsion. We depict the positions after sufficient times of relaxation by dotted circles. While complete reversibility is observed for $\mathcal{V}=0$ in (a), virtually no relaxation takes place for the case of basically absent memory (c). Intermediate amounts of relaxation occur for intermediate strengths of the memory (b).} 

\am{We stress that the depicted dynamical behavior is evaluated by the formalism presented in Sec.~\ref{inclusions}. Accordingly, the mutual interactions mediated between the rigid particles by the viscoelastic environment as well as the finite particle sizes are taken into account to the degree specified above.}

\begin{figure}
\centerline{\includegraphics[width=\columnwidth]{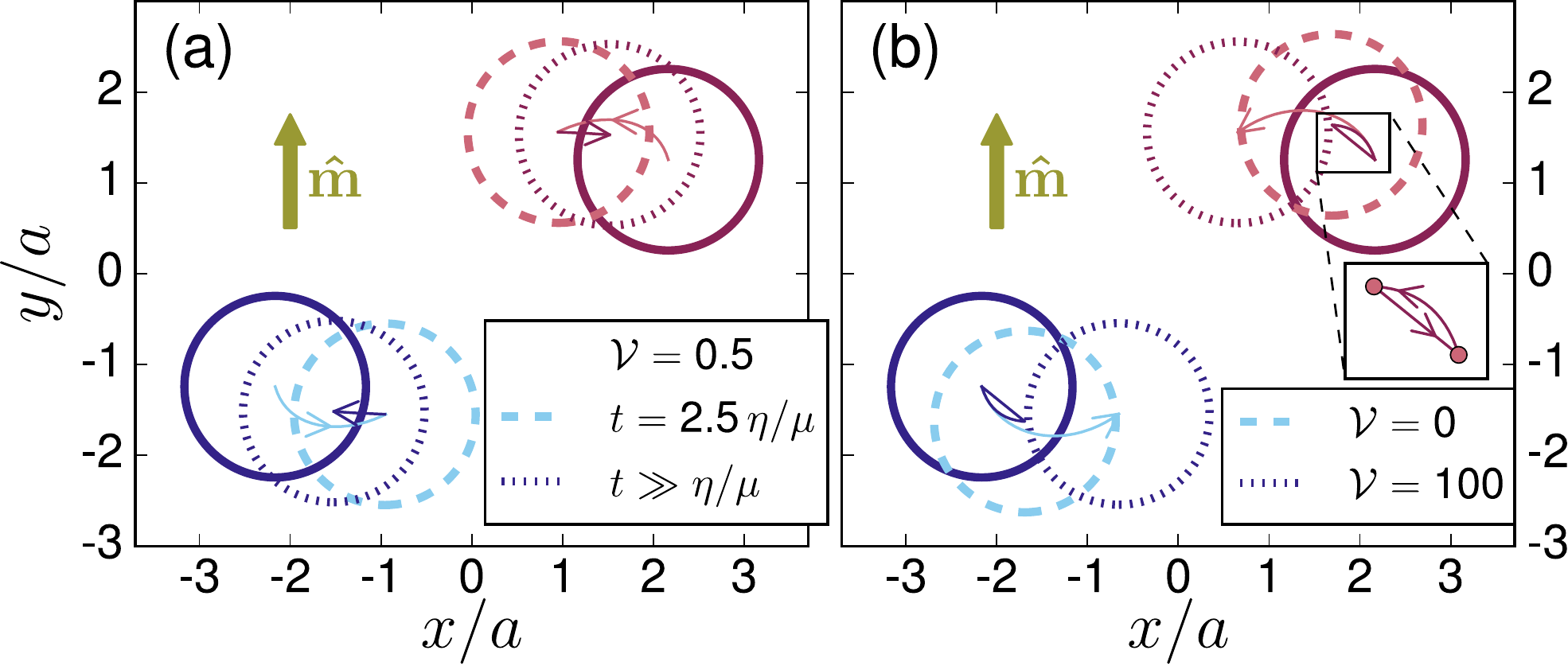}}
\caption{
Same 
as in Fig.~\ref{fig_mag_parallel}, but 
for an oblique configuration and $\mathbf{m} = 200 (\mu a^6 / \mu_0)^{1/2}\mathbf{\hat{m}}$. 
(a) For $\mathcal{V}=0.5$, $\mathbf{m}$ is induced during times $0\leq t\leq 2.5\eta/\mu$. Rearrangements affect the magnetic interactions, while they are constant (zero) for $t>2.5\eta/\mu$. 
Significantly different paths of induced rearrangement (brighter arrows) and 
subsequent relaxation (darker arrows) result. 
(b) The cases $\mathcal{V}=0$ (reversibly elastic solid) and $\mathcal{V}=100$ (nearly memoryless fluid) are shown together for $\mathbf{m}$ induced 
during 
$0\leq t\leq 1.8\eta/\mu$. For $\mathcal{V}=0$, the spheres from their intermediate locations at $t= 1.8\eta/\mu$ (dashed circles) relax back to their initial positions (solid circles) as 
the closed darker trajectories demonstrate. 
For $\mathcal{V}=100$, the spheres first cover longer paths (brighter arrows), but for $t\geq 1.8\eta/\mu$ basically remain in their final locations (dotted circles). 
%
}
\label{fig_mag_versetzt}
\end{figure}

\am{Similar situations, but for initially oblique particle separation vectors relatively to the induced magnetization, are shown in Fig.~\ref{fig_mag_versetzt}.}
\am{Interestingly, in this case the trajectories of initial motion in the magnetized state and of relaxation after the magnetic field has been switched off do not collapse. This even applied to the fully reversible elastic case in Fig.~\ref{fig_mag_versetzt}(b), although the spheres there return to their initial positions. The reason can be associated with the nonreciprocity of the induced magnetic forces during the process. On the forward path, the nonvanishing magnetic forces change during the motion because of their positional dependence. However, they are zero throughout the subsequent relaxational return path.}

\subsection{Active self-propelled microswimmers}

\begin{figure*}
\centerline{\includegraphics[width=2\columnwidth]{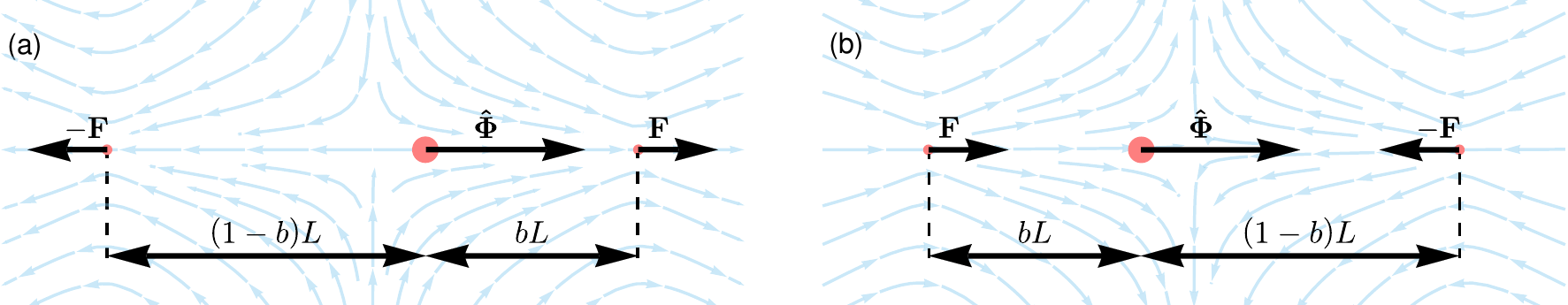}}
\caption{
\am{Illustration of our microswimmer model, adapted from Ref.~\onlinecite{ider2018dynamics}. Two concentrated force centers}, separated by a distance $L$, exert forces $\pm\mathbf{F}=\pm F(t)\mathbf{\hat{\Phi}}(t)$ onto the surrounding medium, set it into motion, and distort it. \am{A sphere 
representing the swimmer body} is asymmetrically placed in between, 
as determined by the parameter $b$,
and is thus subject to net displacements along $\mathbf{\hat{\Phi}}(t)$. The 
resulting minimal model microswimmer displaces and rotates as one rigid entity. A so-called pusher is shown \am{in (a), 
while inverting the forces and thus the flow directions of the background medium transforms it into a puller as depicted in (b). Background arrows indicate the flow field in a viscous fluid.} 
\am{Here, we fix 
$b=0.4$ and $F=\mu L^2$}. 
}
\label{fig_pusher}
\end{figure*}

Another field of significantly growing interest 
concerns 
active self-propelled microswimmers and 
their mutual hydrodynamic interactions 
\cite{saintillan2007orientational,alarcon2013spontaneous,oyama2016purely, elgeti2015physics,bechinger2016active,zottl2016emergent, menzel2015tuned}. 
Increasingly, 
their behavior 
in viscoelastic environments is addressed \cite{gaffney2011mammalian,li2016collective,gomez2016dynamics, lozano2018run, datt2017active,zottl2017enhanced}. 
%
We adapt a recently introduced minimal microswimmer model \cite{menzel2016dynamical,hoell2017dynamical,ider2018dynamics}, see Fig.~\ref{fig_pusher}. 
%
%
%
\am{A spherical swimmer body and two concentrated force centers} are arranged along a common axis oriented 
by the unit vector $\boldsymbol{\hat{\Phi}}(t)$. 
They form one rigid entity that displaces and rotates as one object. 
\am{The spherical swimmer body 
is located at position $\mathbf{R}(t)$, asymmetrically between the two force centers.} 
The latter are separated by a distance $L$, exert axial but oppositely oriented forces $\pm\mathbf{F}(t)=\pm F(t)\mathbf{\hat{\Phi}}(t)$ onto the surrounding medium, thus distorting it and setting it into motion. 
Pushing the medium outward \am{along the symmetry axis as in Fig.~\ref{fig_pusher}(a)} identifies a so-called pusher. Inverting the forces and $\mathbf{\hat{\Phi}}(t)$ 
transforms it into a puller. \am{The latter pulls the medium inward along the symmetry axis, see Fig.~\ref{fig_pusher}(b).} 
\am{This induced} distortion of the surrounding medium leads to a self-induced \am{straight} advective transport \am{of an isolated swimmer} along $\mathbf{\hat{\Phi}}$, \am{if there are no other perturbations. The velocity $\mathbf{V}(t)$ of the swimmer is obtained from the magnitude of the flow field at the location of the swimmer body. That is, $\mathbf{V}(t)=\mathbf{v}(\mathbf{R}(t),t)$, here neglecting the finite extension of the center sphere \cite{ider2018dynamics}.}  
%
%

\begin{figure}
\centerline{\includegraphics[width=\columnwidth]{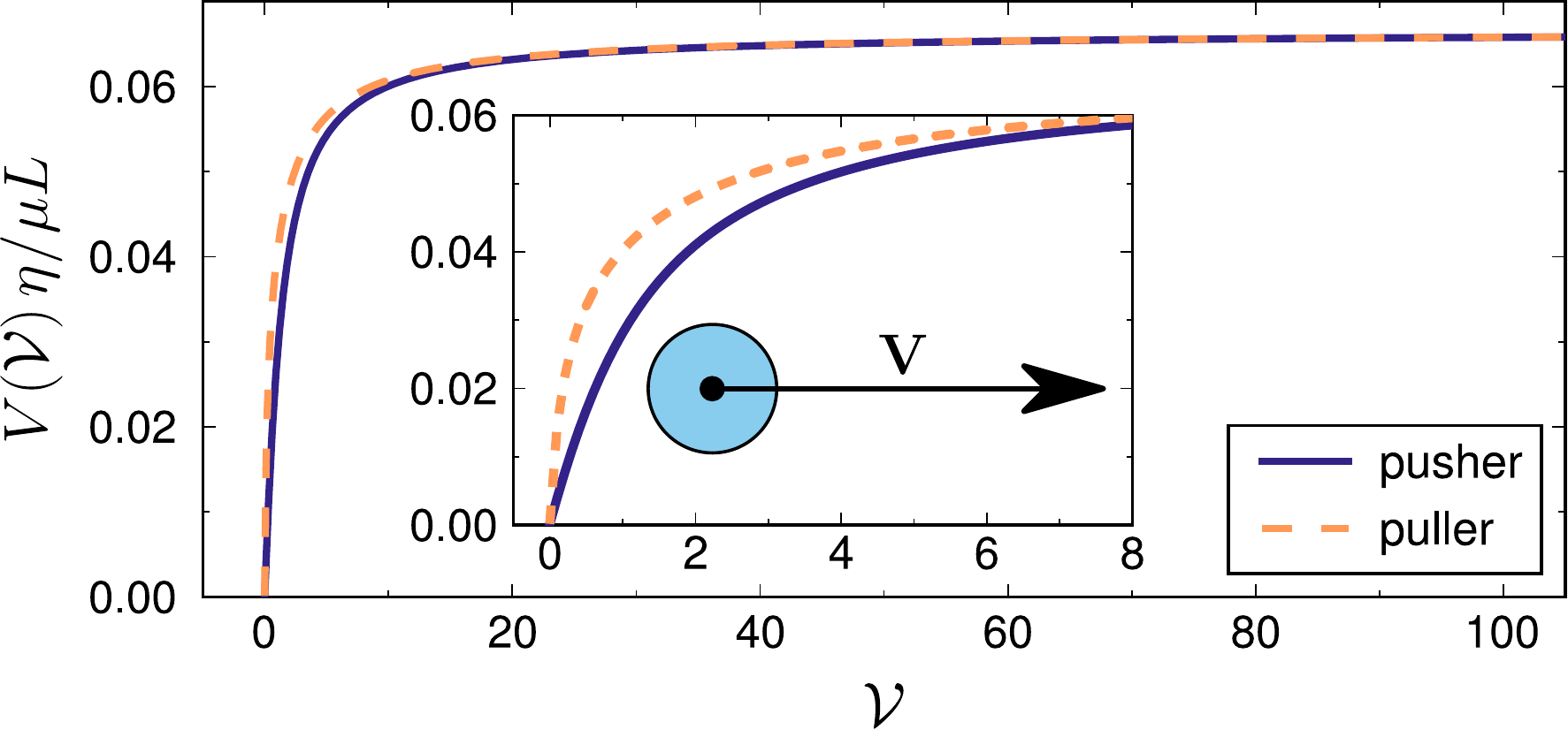}}
\caption{\am{Steady-state swimming speed $V(\mathcal{V})$ as a function of the rescaled ``forgetfulness'' parameter defined in Eq.~(\ref{visco_parameter}). Single, isolated pusher and puller microswimmers are considered as introduced in Fig.~\ref{fig_pusher}. We observe the hydrodynamic value for $\mathcal{V}\gg1$ and a continuous drop with decreasing $\mathcal{V}$ towards $V(\mathcal{V}\rightarrow0)=0$, the latter reflecting the completely reversible elastic case. The inset magnifies the behavior for small values of $\mathcal{V}$.}
}
\label{fig_speed_one_swimmer}
\end{figure}

\am{First, we address individual, isolated microswimmers in the absence of any perturbations. In this context, we analyze how the steady-state swimming speed of the swimmers in Fig.~\ref{fig_pusher} changes with the rescaled ``forgetfulness'' of the viscoelastic environment as defined in Eq.~(\ref{visco_parameter}). Corresponding numerical evaluations can be performed using Mathematica \cite{Mathematica}. Figure~\ref{fig_speed_one_swimmer} shows the steady-state swimming speed $V(\mathcal{V})$ for pushers and for pullers obtained in this way.} 

\am{As may have been expected, for $\mathcal{V}\rightarrow\infty$ the swimming speed tends towards the hydrodynamic value obtained for a microswimmer propelling through a purely viscous fluid. It is identical for pushers and for pullers. In contrast to that, in a completely reversibly deformable elastic environment, the active microswimmer ultimately must come to a rest when the forward drive is balanced by the elastic restoring forces. Thus, $V(\mathcal{V}\rightarrow0)=0$. In between, $V(\mathcal{V})$ continuously drops to zero with decreasing $\mathcal{V}$. That individual pushers in the intermediate regime tend to be slower than individual pullers seems reasonable from Fig.~\ref{fig_pusher}. The swimmer bodies propel into the trace that the heading concentrated force center has left in the viscoelastic medium. Due to the time lag, set by 
the swimming speed $V$, the memorized displacements that were induced on this trace by the heading force center are in the process of relaxation. While the medium around the heading force center of the pusher has been displaced into the swimming direction, it relaxes back into the opposite direction when the swimmer body arrives. Along these lines, a counteracting contribution results for pushers. The reverse follows for pullers.}

\am{In addition to that, we here address the mutual interactions mediated by the viscoelastic environment within a pair of microswimmers that propel alongside each other. Since 
the two microswimmers mutually attract or repel each other, see below, there is no 
steady-state finite distance between the two swimmers. Thus, as a measure for their mutual interaction, we evaluate the flow field alongside one microswimmer in a steady-state motion with velocity $\mathbf{V}$, see Fig.~\ref{fig_speed_one_swimmer}. 
Another microswimmer exposed to this flow field will be affected accordingly. That is, the second swimmer will be sped up or slowed down along its swimming path, and it will be attracted towards or driven away from the first swimmer. Similar arguments apply to the displacements of nearby tracer particles.}

\begin{figure}
\centerline{\includegraphics[width=\columnwidth]{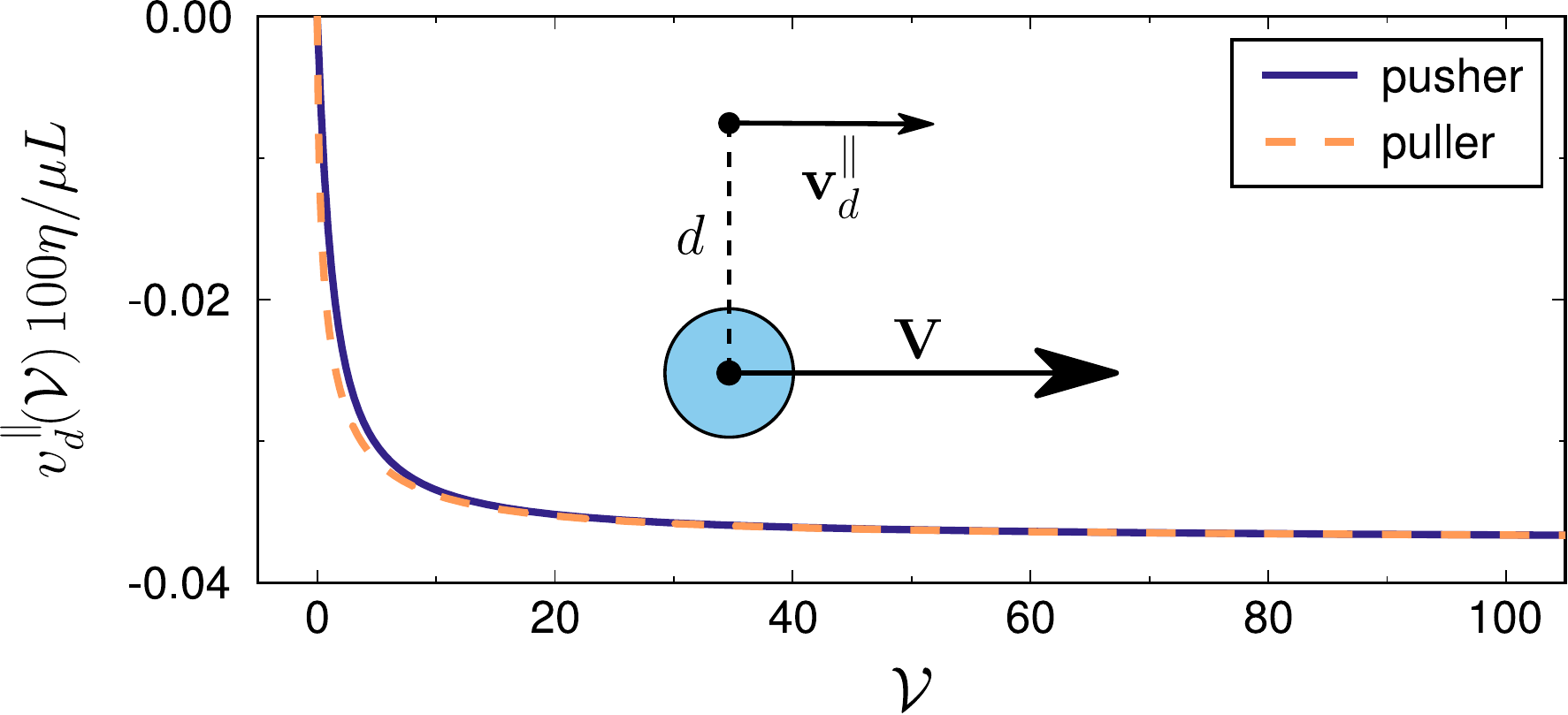}}
\caption{\am{Velocity field $\mathbf{v}_d$ alongside one isolated swimmer in steady-state motion, see Figs.~\ref{fig_pusher} and \ref{fig_speed_one_swimmer}, at a distance $d$ and as a function of the rescaled ``forgetfulness'' parameter $\mathcal{V}$. $\mathbf{v}_d$ is a measure for the mutual interaction between two microswimmers propelling in parallel. Here, the magnitude of the component $\mathbf{v}_d^{\|}$ pointing into the direction of the isolated swimming velocity $\mathbf{V}$ is plotted. $v_d^{\|}<0$ indicates mutual slow-down when compared to the speed of an individual, isolated swimmer in Fig.~\ref{fig_speed_one_swimmer}. We here set $d=2L$.}
}
\label{fig_mutual_par}
\end{figure}

\am{
Accordingly, we evaluate the velocity field $\mathbf{v}_d$ induced in the viscoelastic medium at a distance $d$ alongside the swimmer. If the magnitude of the component $\mathbf{v}_d^{\|}$ pointing into the direction of $\mathbf{V}$ is positive, i.e., $v_d^{\|}>0$, the nearby swimmer propelling in parallel will be sped up. The opposite applies for ${v}_d^{\|}<0$.}
\am{For $d=2L$, corresponding results are depicted in Fig.~\ref{fig_mutual_par}.}
\am{Obviously, since $v_d^{\|}<0$, two swimmers propelling in parallel tend to weakly mutually slow each other down. Figure~\ref{fig_mutual_par} indicates a slightly weaker magnitude of slow-down for pushers than for pullers.} 

\am{The effect of mutual slow-down is in agreement with the flow fields outlined in Fig.~\ref{fig_pusher}. If one continues to draw the induced flow lines 
towards the top and bottom, they will bend away from the vertical center line. Thus, the flow fields there feature a component with direction opposite to the propulsion direction $\mathbf{\hat{\Phi}}(t)$. Swimmers propelling in parallel with their swimmer bodies in such positions will thus mutually slow each other down.}

\begin{figure}
\centerline{\includegraphics[width=\columnwidth]{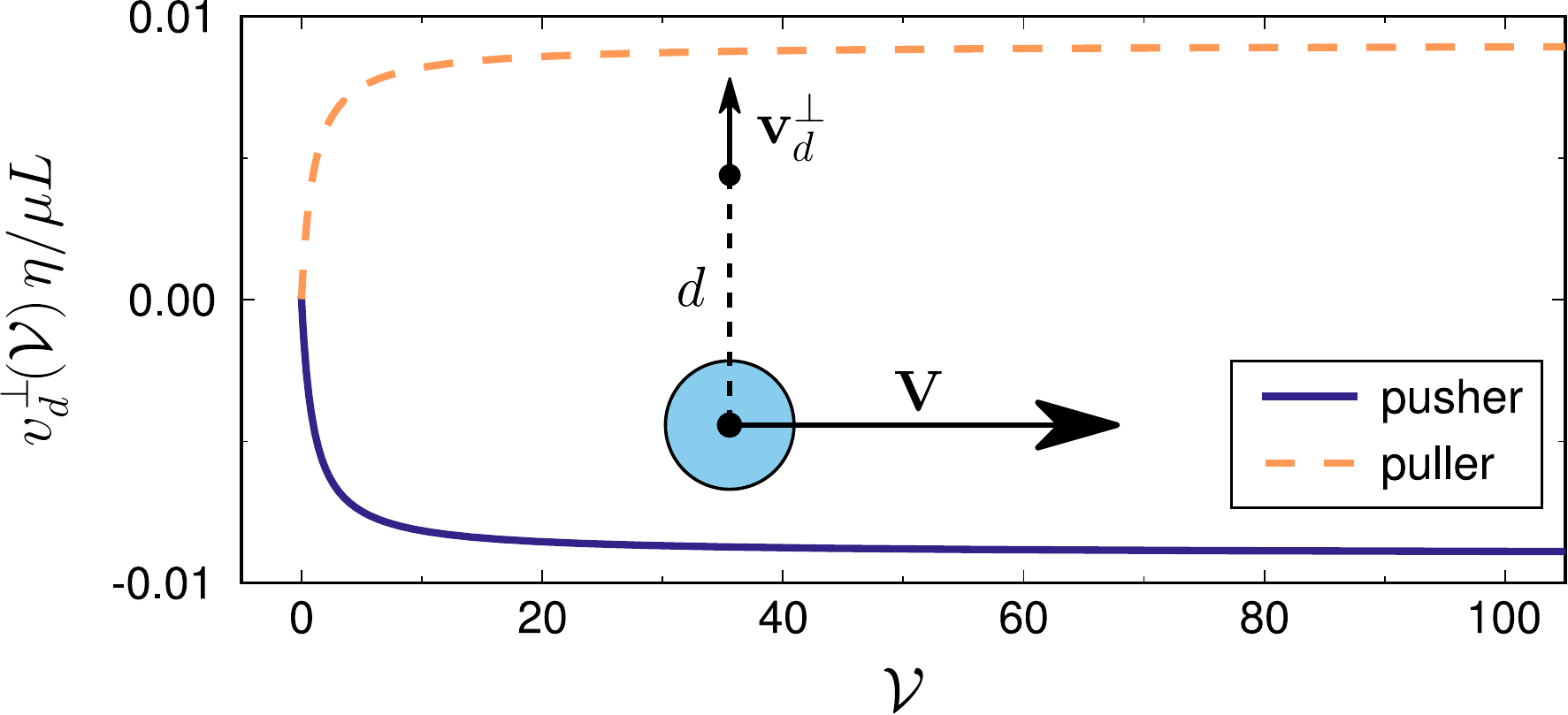}}
\caption{\am{Velocity field $\mathbf{v}_d$ alongside one isolated swimmer in steady-state motion, see Figs.~\ref{fig_pusher} and \ref{fig_speed_one_swimmer}, at a distance $d$ and as a function of the rescaled ``forgetfulness'' parameter $\mathcal{V}$. 
In this case, the magnitude of the component $\mathbf{v}_d^{\bot}$ pointing into a direction perpendicular to $\mathbf{V}$, away from the body of the steadily propelling swimmer, is plotted. $v_d^{\bot}<0$ indicates mutual attraction between two swimmers propelling alongside, while $v_d^{\bot}>0$ marks mutual repulsion. Again, we set $d=2L$.}
}
\label{fig_mutual_perp}
\end{figure}

\am{Similarly, if the magnitude of the component $\mathbf{v}_d^{\bot}$ pointing perpendicularly away from the first swimmer is positive, i.e., $v_d^{\bot}>0$, 
a nearby swimmer propelling alongside will be pushed away, while it will be attracted for ${v}_d^{\bot}<0$. The corresponding results are depicted in Fig.~\ref{fig_mutual_perp}.} 
\am{As can be inferred already from the flow lines in Fig.~\ref{fig_pusher}, pushers attract while pullers repel each other \cite{gotze2010mesoscale}. The effect decreases in magnitude with decreasing $\mathcal{V}$.}

\am{In the future, this approach should be pursued along several different directions. First, the influence of a finite extension of the swimmer body should be investigated, using the formalism outlined in Sec.~\ref{inclusions}. Second, the collective behavior of crowds of active microswimmers in viscoelastic media should be analyzed on the basis of our description, possibly combining it with statistical theories \cite{menzel2016dynamical,hoell2017dynamical}. This might even allow to extend related theories on mesoscale turbulence \cite{heidenreich2016hydrodynamic,reinken2018derivation} to viscoelastic environments.}

\section{Conclusions}\label{conclusions}

In summary, we have introduced and evaluated a formalism 
to describe the dynamic behavior of particulate inclusions in viscoelastic environments when driven by externally imposed, 
induced, 
or actively self-generated forces. 
Interactions 
of the particles 
mediated by the viscoelastic surroundings 
are covered
explicitly. 
Both the limits of a 
reversibly elastic solid-like and a memoryless viscous fluid-like environment are comprised.  

As examples, we characterized the drag of a sphere, the dynamic behavior and mutual interactions of spherical magnetizable particles, 
as well as dynamic properties of 
active microswimmers in viscoelastic media. 
We expect our results to be important, for instance, for the future characterization of the switching dynamics of soft actuation devices \cite{filipcsei2007magnetic, fuhrer2009crosslinking,ilg2013stimuli,thevenot2013magnetic, lee2011programming}, 
to describe the collective dynamics in large ensembles of active microswimmers 
\cite{gaffney2011mammalian,li2016collective,gomez2016dynamics, lozano2018run,yasuda2018three}, to further improve evaluations of active microrheological measurements \cite{ziemann1994local,waigh2005microrheology,wilhelm2008out, khair2010active}, or to extend previous studies on the interactions between living cells in soft environments 
 \cite{schwarz2002elastic,cohen2016elastic}. 
\am{In principle, the procedure can be carried over to other linear models of viscoelasticity as well, to adjust it to the specific properties of a given viscoelastic environment. In the present case, an explicit analytical expression was available in real space for the Green's function and facilitated our evaluation.}


\begin{acknowledgments}
The authors thank G\"unter K.\ Auernhammer and J.~Ruben Gomez-Solano for stimulating discussions, as well as the Deutsche Forschungsgemeinschaft for support of this work through the priority program SPP 1681, grant no.\ ME 3571/3. 
\end{acknowledgments}

\appendix

\section{\am{Continuum description for the viscoelastic background media derived from a generalized hydrodynamic approach}}\label{AppHydro}

\am{Above, the continuum description in Eq.~(\ref{visco}) for the dynamics of the viscoelastic environment was motivated by combining Stokes's equations for the viscous low-Reynolds-number hydrodynamics, Eq.~(\ref{Stokes}), with the Navier--Cauchy equations of linear elasticity, Eq.~(\ref{NC}). Supplementing it by the simple relaxation relation in Eq.~(\ref{velocity_field}), we obtained the continuum characterization in Eq.~(\ref{Combined2}) for the memory displacement field $\mathbf{u}(\mathbf{r},t)$. Here, we demonstrate that these relations can be obtained in a different way as well. For this purpose, we systematically linearize the continuum equations obtained by a previous generalized hydrodynamic theory on viscoelastic materials based on conservation laws and symmetry arguments \cite{temmen2000convective}.} 

\am{In Ref.~\onlinecite{temmen2000convective}, the field $\bm{a}(\mathbf{r},t)$ describes the positions that the material elements currently located at positions $\mathbf{r}$ would tend to take, if all stresses were absent. Thus, $\bm{a}(\mathbf{r},t)=\mathbf{r}-\mathbf{u}(\mathbf{r},t)$. Accordingly, 
the components of the elastic strain tensor 
are given by
\begin{equation}
	U_{ik} ={} \frac{1}{2}\left[\delta_{ik}-(\nabla_i a_l)(\nabla_k a_l)\right],
\end{equation}
with $\delta_{ik}$ the Kronecker delta. 
In the present case, we consider isotropic, homogeneous, infinitely extended viscoelastic environments and therefore do not keep track of reorientations of anisotropy directions of the viscoelastic medium, quantified in Ref.~\onlinecite{temmen2000convective} by the rotation matrix $\mathbf{R}$. Moreover, since in our case the material displacements are assumed to vanish for $|\mathbf{r}|\rightarrow\infty$, we here identify $\bm{a}(\mathbf{r},t)\equiv\mathbf{r}$ for $|\mathbf{r}|\rightarrow\infty$ (and use Latin indices throughout). 
}

\am{Then, our dynamic equation for the motion within the viscoelastic medium is obtained from the dynamic equation in Ref.~\onlinecite{temmen2000convective} for the components of the momentum density $\mathbf{g}(\mathbf{r},t)=\rho(\mathbf{r},t)\mathbf{v}(\mathbf{r},t)$, 
\begin{equation}\label{eq_g}
\dot{g}_i+\nabla_j(\sigma_{ij}-\sigma_{ij}^\text{D})={f}_{b,i},
\end{equation}
where we have added the bulk force density to the right-hand side for our purpose. Assuming incompressibility, the mass density $\rho$ is constant.} 

\am{In Eq.~(\ref{eq_g}), the components of the stress tensor $\sigma_{ij}$ are given by $\sigma_{ij} ={} p\delta_{ij}+v_i g_j +\psi_{l j}\nabla_i a_l$ \cite{temmen2000convective}. Here, $\psi_{l j} = \Psi_{km}(\partial U_{km}/\partial \nabla_j a_l)$ and $\Psi_{km}=\mu U_{km}$. The latter follows via $\mathrm{d}\epsilon=\Psi_{km}\,\mathrm{d}U_{km}$ from the harmonic elastic energy density $\epsilon = K_{ijkm}U_{ij}U_{km}/2$ when expanding $K_{ijkm}=(K_L-K_T/3)\delta_{ij}\delta_{km}+K_T/2(\delta_{ik}\delta_{jm}+\delta_{im}\delta_{jk})$ for isotropic materials \cite{temmen2000convective}, assuming incompressibility, and identifying $\mu\equiv K_T/2$ in our notation.} 

\am{To obtain the appropriate components of the dissipative stresses $\sigma_{ij}^D$, we 
involve the quantity $A_{ij}=(\nabla_i v_j+\nabla_j v_i )/2$ 
\cite{temmen2000convective}, leading to $\sigma_{ij}^D=\eta_{ijkl}A_{kl}$. Expanding the viscosity tensor $\eta_{ijkl}$ similarly to $K_{ijkl}$ above, only one viscosity $\eta$ remains in the end for isotropic incompressible media.}

\am{Finally, we collect all these contributions and insert them into Eq.~(\ref{eq_g}). After strict linearization in $\mathbf{u}(\mathbf{r},t)$ and $\mathbf{v}(\mathbf{r},t)$, neglecting $\dot{g}_i$ in the overdamped situation and for low Reynolds numbers, we indeed obtain our Eq.~(\ref{visco}) from the systematic hydrodynamic approach in Ref.~\onlinecite{temmen2000convective} based on conservation laws and symmetry properties.} 


\am{In view of our dynamic equation for $\mathbf{u}(\mathbf{r},t)$, Eq.~(\ref{velocity_field}), we adopt the relation
\begin{equation}\label{temmen_eq15}
	\dot{U}_{ij} -A_{ij}
	={}-\alpha_T\Psi_{ij}^0-\frac{\delta_{ij}}{3}\alpha_L\Psi_{kk}
\end{equation}
from Ref.~\onlinecite{temmen2000convective}, where we have already dropped obviously nonlinear contributions $\sim v_k\nabla_kU_{ij}$ and $\sim(\nabla_i v_k)U_{kj}$. 
The superscript zero indicates the trace-free part. $\alpha_T$ and $\alpha_L$ denote transport coefficients. 
%
%
Inserting the relations listed above, systematically linearizing in the fields $\mathbf{u}(\mathbf{r},t)$ and $\mathbf{v}(\mathbf{r},t)$, as well as exploiting incompressibility, we obtain
\begin{equation}
	\nabla_i (\dot{u}_j - v_j + 
	2\mu\alpha_Tu_j)
	+\nabla_j (\dot{u}_i - v_i + 
	2\mu\alpha_Tu_i) ={} 0.
\end{equation}
Identifying $\gamma\equiv2\mu\alpha_T$, our Eq.~(\ref{velocity_field}) is in line with this relation, supported by $\mathbf{u}(\mathbf{r},t)=\mathbf{0}=\mathbf{v}(\mathbf{r},t)$ for $|\mathbf{r}|\rightarrow\infty$. 
}

\section{\am{Explicit derivation of the Green's function}}\label{AppGreen} 

Here, we repeat the explicit derivation of the expression for the viscoelastic Green's function $\mathbf{\underline{G}}(\mathbf{r},t)$ in Eq.~(\ref{greens_function_visco}) from the underlying linear partial differential equation in Eq.~(\ref{Combined2}). For this purpose, the incompressibility relations stated in Eqs.~(\ref{Stokes}) and (\ref{NC}) are used. The method works by Fourier transformation both in space and time, solution for the displacement field, and subsequent inverse Fourier transformation. 

Starting from Eq.~(\ref{Combined2}), the Green's function $\mathbf{\underline{G}}(\mathbf{r},t)$ quantifies the displacement field $\mathbf{u}(\mathbf{r},t)$ created by a point force impact $\mathbf{F}$ resulting from 
\begin{equation}\label{force_fb}
\mathbf{f}_b(\mathbf{r},t)=\mathbf{F}\delta(\mathbf{r}-\mathbf{R}_0)\delta(t-t_0),
\end{equation}
where $\mathbf{R}_0$ and $t_0$ set the position and time of attack, respectively.
That is, the resulting displacement field reads $\mathbf{u}(\mathbf{r},t)=\mathbf{\underline{G}}(\mathbf{r}-\mathbf{R}_0,t-t_0)\cdot\mathbf{F}$. 
Without loss of generality, we choose $\mathbf{R}_0=\mathbf{0}$ and $t_0=0$, for simplicity, so that this relation becomes
\begin{equation}\label{def_G}
\mathbf{u}(\mathbf{r},t)=\mathbf{\underline{G}}(\mathbf{r},t)\cdot\mathbf{F}.
\end{equation}
%
%
%

Integrating both sides of Eq.~(\ref{Combined2}) over
\begin{equation}\label{fourier_transform}
	\frac{1}{(2\pi)^2}\int_{\mathbb{R}^3}\mathrm{d}^3r~e^{-i\mathbf{k}\cdot\mathbf{r}}\int_\mathbb{R}\mathrm{d}t~e^{-i\omega t},
\end{equation}
we obtain its space-and-time Fourier transform as
\begin{equation}\label{eq_motion_fourier}
	k^2(\mu+\gamma\eta+i\omega\eta) \mathbf{\tilde{u}}(\mathbf{k},\omega) ={} - i \mathbf{k}\tilde{p}(\mathbf{k},\omega) + \mathbf{\tilde{f}}_b(\mathbf{k},\omega),
\end{equation}
with Fourier-transformed quantities marked by the tilde. Equation~(\ref{force_fb}) implies
\begin{equation}
	\mathbf{\tilde{f}}_b(\mathbf{k}, \omega) ={}  \frac{1}{(2\pi)^2}\mathbf{F}.
\end{equation}

Next, we involve the incompressibility relations stated in Eqs.~(\ref{Stokes}) and (\ref{NC}). Together with Eq.~(\ref{velocity_field}), they read
\begin{equation}\label{incomp}
	\nabla\cdot\mathbf{{{u}}}(\mathbf{r}, t) ={} 0, \qquad \nabla\cdot\mathbf{{\dot{u}}}(\mathbf{r}, t) ={} 0.
\end{equation}
Their space-and-time Fourier transforms via Eq.~(\ref{fourier_transform}) follow as
\begin{equation}\label{Incompressibility}
	\mathbf{k}\cdot\mathbf{\tilde{u}}(\mathbf{k}, \omega) ={} 0, \qquad \mathbf{k}\cdot\mathbf{\tilde{\dot{u}}}(\mathbf{k}, \omega) ={} 0,
\end{equation}
from which the relations $\mathbf{k}\perp\mathbf{\tilde{\dot{u}}}(\mathbf{k}, \omega)$ and $\mathbf{k}\perp\mathbf{\tilde{u}}(\mathbf{k}, \omega)$ are obtained.
Multiplication of Eq.~(\ref{eq_motion_fourier}) by the projection operator $\mathbf{\underline{\hat{I}}}-\mathbf{\hat{k}}\mathbf{\hat{k}}$, with $\mathbf{\hat{k}}=\mathbf{k}/|\mathbf{k}|$ and $\mathbf{\underline{\hat{I}}}$ denoting the unity matrix, then solving for $\mathbf{\tilde{u}}$ yields
\begin{equation}
	\mathbf{\tilde{u}}(\mathbf{k}, \omega) ={} \frac{1}{k^2(\mu^\star+i\omega\eta)}(\mathbf{\underline{\hat{I}}}-\mathbf{\hat{k}}\mathbf{\hat{k}})\cdot\frac{1}{(2\pi)^2}\mathbf{F}	\label{u-tilde},
\end{equation}
where $\mu^\star:=\mu+\gamma\eta$.
We can rewrite Eq.~(\ref{u-tilde}) as
\begin{equation}
	\mathbf{\tilde{u}}(\mathbf{k}, \omega) ={} \mathbf{\underline{\hspace{-.02cm}\tilde{G}}}(\mathbf{k}, \omega) \cdot \mathbf{F},
\end{equation}
which corresponds to the Fourier transform of Eq.~(\ref{def_G}). Therefore, 
\begin{equation}\label{G_FT}
	\mathbf{\underline{\hspace{-.02cm}\tilde{G}}}(\mathbf{k}, \omega) ={} \frac{1}{(2\pi)^2 k^2(\mu^\star + i \omega \eta)} (\mathbf{\hspace{.02cm}\underline{\hat{I}}} - \mathbf{\hat{k}}\mathbf{\hat{k}})
\end{equation}
is the space-and-time Fourier transform of the Green's function.
To obtain the Green's function in real space, we then need to calculate the inverse Fourier transformation,
\begin{equation}
	\mathbf{\hspace{.02cm}\underline{\hspace{-.02cm}G}}(\mathbf{r}, t) 
	={}  \frac{1}{(2\pi)^2} \int_{\mathbb{R}^3} \mathrm{d}^3k~e^{i\mathbf{k}\cdot\mathbf{r}}\int_{\mathbb{R}} \mathrm{d}\omega~e^{i\omega t}\mathbf{\tilde{G}}(\mathbf{k}, \omega).
\end{equation}

First, we evaluate the $\int \mathrm{d}\omega$ integral,
\begin{equation}
	\frac{1}{\sqrt{2\pi}i\eta} \int_\mathbb{R}\mathrm{d}\omega \frac{e^{i\omega t}}{\omega-i \frac{\mu^\star}{\eta}},
\end{equation}
with a singularity in the complex plane at $\omega = i \mu^\star/\eta$.
For $t>0$, the integration path is closed in the upper half plane.
For $t<0$, the integration path is closed in the lower half plane, where there is no singularity.
Using the residue theorem, we thus find
%
%
\begin{equation}
	\frac{1}{\sqrt{2\pi}i\eta} \int_\mathbb{R}\mathrm{d}\omega \frac{e^{i\omega t}}{\omega-i \frac{\mu^\star}{\eta}} ={} \frac{\sqrt{2\pi}}{\eta} \Theta(t) e^{-\frac{\mu^\star}{\eta}t}, 
\end{equation}
where $\Theta(\cdot)$ denotes the Heaviside step function. 

Then, the inverse transformation in space remains as
\begin{equation}
	\mathbf{\underline{G}}(\mathbf{r}, t) ={} C \int_{\mathbb{R}^3} \mathrm{d}^3k \frac{e^{i\mathbf{k}\cdot\mathbf{r}}}{k^2}(\mathbf{\hspace{.02cm}\underline{\hat{I}}} - \mathbf{\hat{k}}\mathbf{\hat{k}}),
\end{equation}
with the abbreviation
\begin{equation}
	C ={} \frac{\Theta(t)}{(2\pi)^3\eta} e^{-\frac{\mu^\star}{\eta}t}.
\end{equation}
As an ansatz for $\mathbf{\underline{G}}(\mathbf{r},t)$, we choose \cite{doi1988theory}
\begin{equation}
	\mathbf{\hspace{.02cm}\underline{\hspace{-.02cm}G}}(\mathbf{r},t) ={} A\mathbf{\hspace{.02cm}\underline{\hat{I}}} + B\mathbf{\hat{r}}\mathbf{\hat{r}},
\end{equation}
implying $A=A(r,t)$, $B=B(r,t)$, $r=|\mathbf{r}|$, and $\mathbf{\hat{r}}=\mathbf{r}/r$. 
First, from the trace $G_{jj}(\mathbf{r},t)$, where summation over repeated indices is implied, we obtain the relation
\begin{eqnarray}
	3A+B &={} & 2 C \int\limits_{0}^{2\pi}\mathrm{d}\varphi\int\limits_{-1}^{1}\mathrm{d}\hspace{-.15em}\cos\vartheta\int\limits_{0}^{\infty}\mathrm{d}k~e^{ikr\cos\vartheta}\notag\\
	&={} & \frac{8\pi C}{r} \int\limits_{0}^{\infty} \mathrm{d}k~\frac{\sin(kr)}{k}\notag\\
		&={} & \frac{8\pi C}{r} \int\limits_{0}^{\infty} \mathrm{d}\xi~ \frac{\sin\xi}{\xi}\notag\\
		&={} & \frac{4\pi^2 C}{r}. \label{appendix_3a_b}
\end{eqnarray}
Analogously, the contraction $G_{ij}(\mathbf{r},t)\hat{r}_i\hat{r}_j$ yields the relation
\begin{eqnarray}
	A+B &={} & 2\pi C \int\limits_{-1}^{1}\mathrm{d}\hspace{-.15em}\cos\vartheta\int\limits_{0}^{\infty}\mathrm{d}k~  (1-\cos^2\vartheta)  e^{ikr\cos\vartheta} \notag\\
	&={} & \frac{2\pi C}{r} \int\limits_{0}^{\infty} \mathrm{d}\xi \int\limits_{-1}^{1}\mathrm{d}\hspace{-.15em}\cos\vartheta \left(1+\frac{\partial^2}{\partial \xi^2}\right) e^{i\xi\cos\vartheta} \notag\\
	&={} & \frac{4\pi C}{r} \int\limits_{0}^{\infty} \mathrm{d}\xi \left(1+\frac{\partial^2}{\partial \xi^2}\right) \frac{\sin\xi}{\xi} \notag\\
	&={} & \frac{2\pi^2 C}{r}. \label{appendix_a_b}
\end{eqnarray}
In combination, we find from Eqs.~(\ref{appendix_3a_b}) and (\ref{appendix_a_b}) that
\begin{equation}
	A ={} B={} \frac{\Theta(t)}{8\pi\eta r}e^{-\frac{\mu^\star}{\eta}t},
\end{equation}
resulting in Eq.~(\ref{greens_function_visco}) for the viscoelastic Green's function.

\section{\am{Kramers--Kronig relations}}\label{AppKramers}

Next, we briefly demonstrate that the Green's function derived in Eqs.~(\ref{greens_function_visco}) and (\ref{G_FT}) indeed satisfies the famous Kramers--Kronig relations \cite{jackson1962classical}. 
To this end, we solve Eq.~(\ref{eq_motion_fourier}) for $\mathbf{\tilde{u}}(\mathbf{k},\omega)$, and apply to both sides of the equation the projection operator 
$\mathbf{\hspace{.02cm}\underline{\hat{I}}}-\mathbf{\hat{k}}\mathbf{\hat{k}}$, 
see above.
Exploiting Eq.~(\ref{Incompressibility}), this projection leaves $\mathbf{\tilde{u}}(\mathbf{k},\omega)$ unchanged.
From the resulting equation of the form
\begin{equation}
\mathbf{\tilde{u}}(\mathbf{k},\omega)=\chi(k,\omega)(\mathbf{\underline{\hat{I}}}-\mathbf{\hat{k}}\mathbf{\hat{k}})\cdot\mathbf{\tilde{f}}_b(\mathbf{k},\omega),
\end{equation}
we thus obtain an effective susceptibility
\begin{equation}\label{susceptability}
	\chi(k,\omega) ={} \frac{1}{k^2}\frac{\mu+\gamma\eta -i \omega\eta}{(\mu+\gamma\eta)^2 +\omega^2\eta^2}.
\end{equation}
It has a shape related to the Kelvin--Voigt model \am{[although the additional parameter $\gamma$ plays a decisive qualitative role in the present case and controls the long-term system behavior; this becomes clear from the rescaling at the beginning of Sec.~\ref{draggedsphere}, which identifies $\mathcal{V}=\gamma\eta/\mu$ in Eq.~(\ref{visco_parameter}) as the one remaining parameter to quantify the behavior of the viscoelastic environment]}.
We can rewrite $\chi({k},\omega)$ as
\begin{equation}
	\chi(k,\omega) ={} \chi_1(k,\omega) + i\chi_2(k,\omega),
\end{equation}
with $\chi_1(k,\omega)=\Re\chi(k,\omega)$ an even and $\chi_2(k,\omega)=\Im\chi(k,\omega)$ an odd function of $\omega$.
A singularity is located in the upper complex half-plane at $\omega=i(\mu+\gamma\eta)/\eta$.
Then, the Kramers--Kronig relations for positive frequencies $\omega$ can be formulated as
\begin{eqnarray}
	\chi_1(k,\omega) &={} &{} -\frac{2}{\pi}\mathcal{P}\int\limits_{0}^{\infty}\mathrm{d}\omega'~\frac{\omega'\chi_2(k,\omega')}{\omega'^2-\omega^2}, \label{kramers_1}\\
	\chi_2(k,\omega) &={} &{} \frac{2\omega}{\pi}\mathcal{P}\int\limits_{0}^{\infty}\mathrm{d}\omega'~\frac{\chi_1(k,\omega')}{\omega'^2-\omega^2}, \label{kramers_2}
\end{eqnarray}
linking the real and imaginary parts of $\chi(k,\omega)$ to each other.
Since $\chi(k,\omega)$ vanishes as $\sim\omega^{-1}$ for $\omega\rightarrow\infty$ and $\eta\neq0$, the requirements for the Kramers--Kronig relations Eqs.~(\ref{kramers_1}) and (\ref{kramers_2}) to be satisfied are met.

\section{\am{Basic example trajectory of a material element subject to an interim constant concentrated force density}}\label{AppMatEl}

\am{In this slightly academic example, we further illustrate via analytical formulae the background of our description introduced in Eqs.~(\ref{visco}) and (\ref{velocity_field}). For this purpose, we consider the force density, 
\begin{equation}
\mathbf{f}_b(\mathbf{r},t)=F\mathbf{\hat{x}}\,\delta(\mathbf{r})\big(\Theta(t)-\Theta(t-t_e)\big),
\end{equation}
spatially concentrated at the origin, pointing with constant magnitude $F>0$ into the $x$ direction, switched on at time $t=0$, and turned off at time $t=t_e$. Moreover, we here only consider positions $\mathbf{r}$ on the positive $x$ axis, parameterized by $\mathbf{r}=r\mathbf{\hat{x}}$, with $r>0$.  Consequently, the spatial integral in Eq.~(\ref{displacement-force-field}) reduces to 
\begin{equation}
\int_{\mathbb{R}^3}\! \mathrm{d}^3r'~
	 \mathbf{\underline{G}}(\mathbf{r}-\mathbf{r}')\cdot\mathbf{f}_b(\mathbf{r}',t')
	 = \frac{F}{4\pi\eta r}\mathbf{\hat{x}}
  	 \big(\Theta(t')-\Theta(t'-t_e)\big). 
\end{equation}
Obviously, for $t<0$, there is no motion and displacement of material elements.} 

\am{For $0<t<t_e$, via Eqs.~(\ref{velocity_field}) and (\ref{displacement-force-field}), we calculate the velocity on the positive $x$ axis as 
\begin{equation}
\mathbf{v}(\mathbf{r},t)=\frac{F}{4\pi\eta r}\mathbf{\hat{x}}\left(
1-\frac{1}{1+\mathcal{V}}\left(1-\mathrm{e}^{-\frac{\mu+\gamma\eta}{\eta}t}\right)\right),
\end{equation}
where we have used the definition in Eq.~(\ref{visco_parameter}). The trajectory $\mathbf{R}(t)$ of a material element initially located on the positive $x$ axis at position $\mathbf{R}(t<0)=R_0\mathbf{\hat{x}}$, with $R_0>0$, therefore remains confined to the $x$ axis. From
\begin{equation}\label{eq_R_v}
\frac{\mathrm{d}\mathbf{R}(t)}{\mathrm{d}t}=\mathbf{v}(\mathbf{R}(t),t),
\end{equation}
with $\mathbf{R}(t)=R(t)\mathbf{\hat{x}}$, we thus obtain
\begin{equation}\label{R_at_te}
R(t)^2=R_0^2+\frac{F}{2\pi\eta}\frac{1}{1+\mathcal{V}}\left(\!
\mathcal{V}t+\frac{\eta}{\mu+\gamma\eta}\left(1-\mathrm{e}^{-\frac{\mu+\gamma\eta}{\eta}t}\right)\!\right). 
\end{equation}}

\am{Afterwards, for $t>t_e$, the velocity field on the positive $x$ axis becomes
\begin{equation}
\mathbf{v}(\mathbf{r},t)={}-\frac{F}{4\pi\eta r}\mathbf{\hat{x}}\frac{1}{1+\mathcal{V}}\left(\mathrm{e}^{\frac{\mu+\gamma\eta}{\eta}t_e}-1\right)\mathrm{e}^{-\frac{\mu+\gamma\eta}{\eta}t}.
\end{equation}
Consequently, the trajectory is determined by 
\begin{eqnarray}
R(t)^2&=&R_0^2 +\frac{F}{2\pi\eta}\frac{1}{1+\mathcal{V}}
\bigg(\mathcal{V}t_e
\nonumber\\
&&{}
+\frac{\eta}{\mu+\gamma\eta}\left(\mathrm{e}^{\frac{\mu+\gamma\eta}{\eta}t_e}-1\right)\mathrm{e}^{-\frac{\mu+\gamma\eta}{\eta}t}\bigg),
\quad
\label{trajectory}
\end{eqnarray}
which follows again via Eq.~(\ref{eq_R_v}) and upon inserting Eq.~(\ref{R_at_te}) for $t=t_e$.} 

\am{If we measure time in units of $\eta/\mu$ and again use the definition in Eq.~(\ref{visco_parameter}), then Eq.~(\ref{trajectory}) becomes
\begin{eqnarray}
R(t)^2&=&R_0^2 +\frac{F}{2\pi\mu}\frac{1}{1+\mathcal{V}}
\bigg(\mathcal{V}t_e
\nonumber\\
&&{}
+\frac{1}{1+\mathcal{V}}\left(\mathrm{e}^{(1+\mathcal{V})t_e}-1\right)\mathrm{e}^{-(1+\mathcal{V})t}\bigg).
\quad
\label{erg_Rt}
\end{eqnarray}
We can infer from this basic example the relevance of the parameter $\mathcal{V}$. In the fully reversible elastic situation, i.e., for $\mathcal{V}=0$, the material element at long times $t\rightarrow\infty$ returns to its initial position $\mathbf{R}(t=0)=R_0\mathbf{\hat{x}}$. This is not the case for $\mathcal{V}\neq0$. Thus, the parameter $\mathcal{V}$ qualitatively determines the behavior of the system. One could perform another rescaling of time to remove the factor $1+\mathcal{V}$ from the exponents, but this does not eliminate the influence of $\mathcal{V}$ on the first contribution of $t_e$ in Eq.~(\ref{erg_Rt}).}

\section{\am{Additional net torques applied to the spherical particles}}\label{AppTorque}

The effect of net forces acting on the spherical particles has been addressed in Eq.~(\ref{u_general}). Since the underlying equations are linear in the displacement field, see Eq.~(\ref{Combined2}), the effects of additional torques can simply be superimposed. 

We proceed in the same way as in the case of the applied forces. Applying a net torque $\mathbf{T}(t)$ to a spherical particle of radius $a$ 
in analogy to the case of an applied force, we consider the displacement field 
\begin{eqnarray}
	\mathbf{u}(\mathbf{r},t)
		&=&{} \int_\mathbb{R}\mathrm{d}t'~G(t-t')\:
\Bigg\{		
    \nonumber\\ &&{}
		-\frac{1}{2}\,\mathbf{T}(t')\cdot\left[\nabla\times
		\mathbf{\underline{G}}\big(\mathbf{s}(t')\big)\right]
		\Theta\big(|\mathbf{s}(t')|-a
				\big)	
		\nonumber\\ &&{}
		+\frac{1}{8\pi\eta a^3}\,\mathbf{T}(t')\times \mathbf{s}(t')~\Theta\big(a-|\mathbf{s}(t')|
						\big)
		\Bigg\},\qquad
		\label{torque}
\end{eqnarray}
%
%
%
%
with 
$\mathbf{s}(t'):=\mathbf{r}-\mathbf{R}(t')$ and again a continuous integrand for each $|\mathbf{s}(t')|=a$. 
In analogy to Eq.~(\ref{u_general}), the expression in Eq.~(\ref{torque}) solves for $|\mathbf{s}(t)|>a$, i.e., inside the embedding medium, 
the linear Eq.~(\ref{Combined2}) for $\mathbf{f}_b(\mathbf{r},t)=\mathbf{0}$, together with $\nabla\cdot\mathbf{u}(\mathbf{r},t)=0$.  Moreover, $\mathbf{u}(\mathbf{r},t)\rightarrow\mathbf{0}$ for $|\mathbf{s}(t)|\rightarrow\infty$, see Eq.~(\ref{greens_function_visco}), if all $|\mathbf{R}(t')|$ remain finite. 
Additionally, at each instant in time $t'$, the expression in square brackets now leads to a rigid rotation of all points on the surface of the sphere, 
i.e., for $|\mathbf{s}(t')|=a$. 
The latter can be shown by explicitly evaluating $\nabla\times\mathbf{\underline{G}}\big(\mathbf{r}-\mathbf{R}(t')\big)$.
As for the case of applied forces in Eq.~(\ref{u_general}), the displacements and rotations of the sphere at time $t$ resulting from the contributions generated in Eq.~(\ref{torque}) at earlier times $t'<t$ 
can then be calculated via Eqs.~(\ref{u_integral}) and (\ref{omega_integral}).

\end{document}